\documentclass[english,aps,prl,twocolumn,superscriptaddress,showpacs]{revtex4}
\usepackage[T1]{fontenc}
\usepackage[latin9]{inputenc}
\usepackage{babel}

\usepackage{graphicx}
\usepackage{amssymb}

\begin{document}

\title{Creating p-wave superfluids and topological excitations in optical lattices}

\author{P. Massignan}

\affiliation{ICFO-Institut de Ci\`encies Fot\`oniques, Mediterranean Technology Park,
08860 Castelldefels (Barcelona), Spain.}

\affiliation{Grup de F\'isica Te\'orica, Universitat Aut\`onoma de Barcelona, 08193
Bellaterra, Spain }

\author{A. Sanpera}

\affiliation{Grup de F\'isica Te\'orica, Universitat Aut\`onoma de Barcelona, 08193
Bellaterra, Spain }

\affiliation{ICREA-Instituci\'o Catalana de Recerca i Estudis Avan\c{c}ats, 08010 Barcelona,
Spain.}

\author{M. Lewenstein}

\affiliation{ICFO-Institut de Ci\`encies Fot\`oniques, Mediterranean Technology Park,
08860 Castelldefels (Barcelona), Spain.}

\affiliation{ICREA-Instituci\'o Catalana de Recerca i Estudis Avan\c{c}ats, 08010 Barcelona,
Spain.}


\date{\today}
\begin{abstract}
We propose to realize a p-wave superfluid using bosons mixed with a single species of fermions in a deep optical lattice. We analyze with a self-consistent method its excitation spectrum in presence of a vortex, and we point out the range of interaction strengths in which the zero-energy mode with topological character exists on a finite optical lattice. Lattice effects are strongest close to fermionic half-filling: here the linearity of the low-lying spectrum is lost, and a new class of extended zero-energy modes with checkerboard structure and d-wave symmetry appears.

\end{abstract}
\maketitle

Pairing physics plays a central role in systems as diverse as neutron stars, liquid Helium, solid state superconductors and 
ultracold dilute gases \cite{Leggett06}. Important advances in our understanding of this fascinating phenomenon 
came from the recent theoretical and experimental progresses made with two-component fermionic gases \cite{Bloch2008,Giorgini08}. 
In these systems, pairing occurs mainly in the s-wave channel and pairs have vanishing angular momentum.
Ultracold atoms may provide the highly-controlled experimental framework needed to prepare and study unconventional superfluids, composed of pairs having a non-zero angular momentum. In two dimensions (2D), these superfluids feature anyonic excitations with a linear (Dirac-like) spectrum, and zero-energy excitations with topological character and non-Abelian statistics \cite{Read00,Tewari07,Gurarie07, Bergman09}. For p-wave symmetry, these excitations are analogous to the ones found in the $\nu=5/2$ Fractional Quantum Hall physics.
 Due to their insensibility to local perturbations, non-Abelian anyons are very appealing for applications in Topologically-protected Quantum \mbox{Computation \cite{Nayak08}}.

 Although strong interactions between identical fermions have been obtained close to a p-wave Feshbach resonance \cite{Chin00,Regal03,Zhang04}, experiments so far failed to reach the superfluid regime since three-body collisions severely limit the lifetime of the gas. 
 Recent theoretical proposals to realize long lived p-wave superfluids relied either on ensembles of 2D fermionic polar molecules dressed with radio-frequency \cite{Cooper09} or on suppression of three-body collisions through the Quantum Zeno effect \cite{Han09}.
Realization of p-wave and d-wave superfluids in shallow lattices has been proposed in Refs.~\cite{Wang05,Lim09}. Large p-wave interactions are also predicted to arise between 3D lattice fermions in contact with 2D dipolar bosons \cite{Dutta09}.

\begin{figure}
\includegraphics[width=0.33\columnwidth]{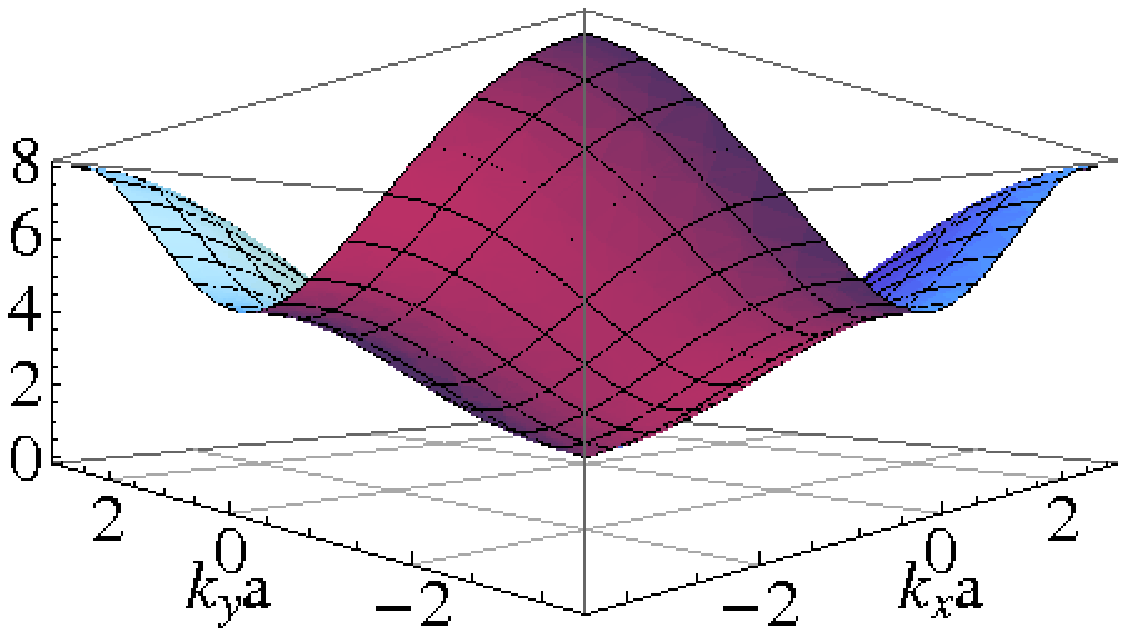}\includegraphics[width=0.33\columnwidth]{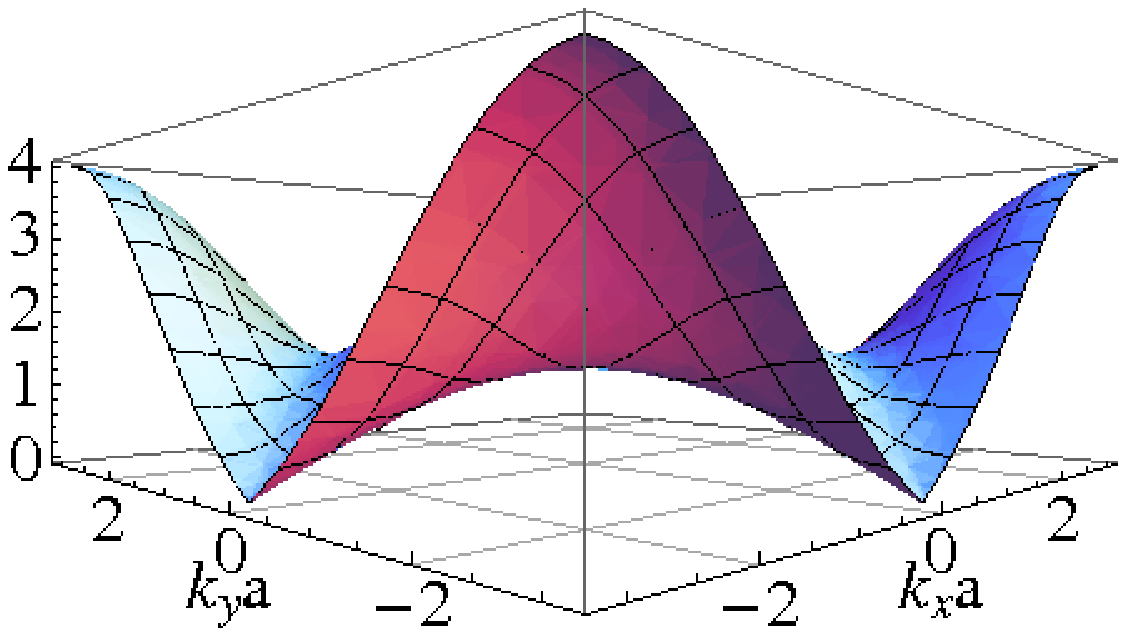}\includegraphics[width=0.33\columnwidth]{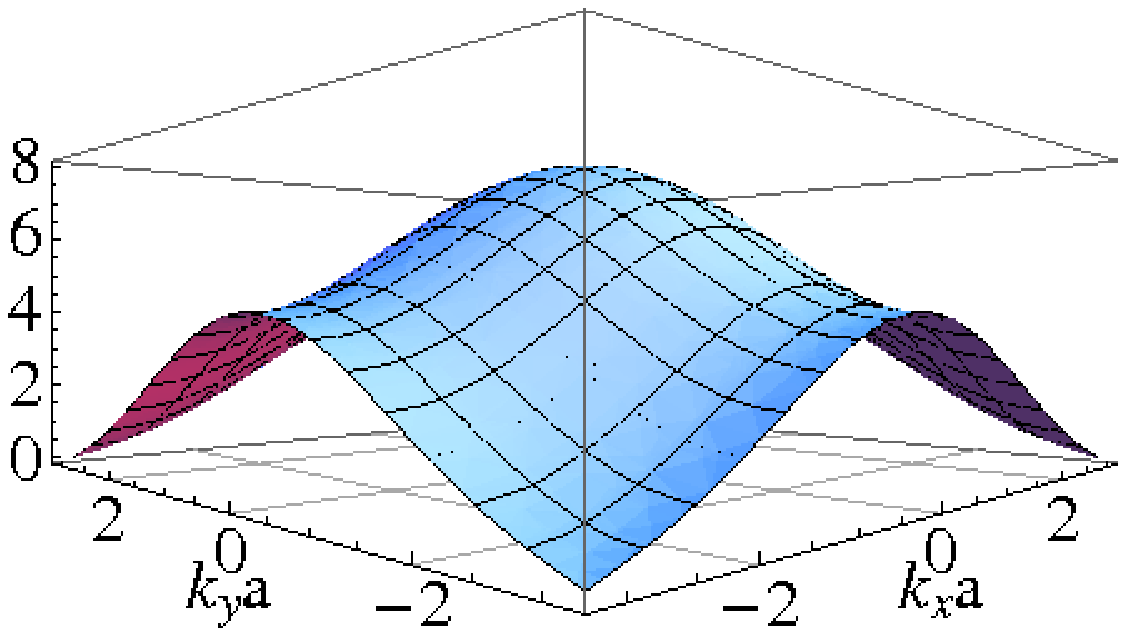}
\includegraphics[width=\columnwidth]{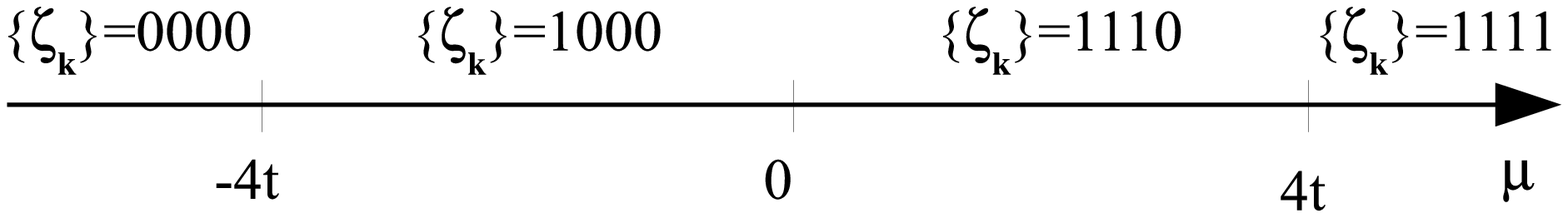}

\caption{The dispersion $E(\mathbf{k})/t$ of a vortex-free chiral p-wave
superfluid on a lattice, Eq.~(\ref{eq:dispersionRel}), vanishes linearly around the Dirac points. These gapless points appear, from left to right, at $\mu=-4t,0,4t$, and they separate different topological phases, each identified by a set $\{\zeta_{\bf k}\}$. The four indices of the set are 0 or 1, depending on whether the momentum state with $\textbf{k}d=\{(0,0),(\pi ,0),(0,\pi),(\pi,\pi)\}$ is empty or occupied \cite{Kou09}. While the topological transition at $\mu=-4t$ is analogous to the one found in the continuum, the ones at $\mu=0,4t$ are entirely due to lattice physics.
\label{fig:dispersionRel}}
\end{figure}

In this paper, we propose to realize a stable p-wave superfluid in an optical lattice by doping a bosonic Mott state with a \emph{single} fermionic species. We study its excitation spectrum in presence of a vortex by means of a self-consistent Bogolubov-de Gennes (BdG) approach. As shown in the seminal paper of Read and Green \cite{Read00}, a 2D p-wave superfluid in the continuum has a zero energy excitation (Majorana mode) with exotic topological properties. In a lattice of finite extension, comparable to the ones found in current experiments, we find that the zero-energy Majorana mode survives only for intermediate interaction strengths. Moreover, the lattice brings along a number of peculiar effects. In particular, we find that at half filling the spectrum becomes gapless and the gas undergoes a Topological Phase Transition (TPT). Close to criticality, a new class of extended zero-energy modes appears, characterized by a distinctive checkerboard structure and d-wave symmetry.



Our proposal to form a p-wave gas employs a Bose-Fermi mixture in a deep lattice. We choose the chemical potential for the bosons such that, in the absence of fermions, the bosons form a Mott state with unit filling. As described in Ref.~\cite{Lewenstein04}, adding identical fermions with filling $0<F<1$ yields a system that may be described by an effective Hamiltonian for composite fermions (fermion+boson, or fermion+bosonic hole, depending on the sign of the interspecies interaction). At low temperatures the identical composite fermions cannot collide in s-wave due to Pauli exclusion principle, and the interaction has a dominant p-wave character. Superfluidity  occurs when both Bose-Fermi and Bose-Bose interactions are repulsive and satisfy $|U_{bf}|>|U_{bb}|>0$, giving an attractive (superexchange) nearest-neighbor interaction between fermion-hole composites; this interaction is characterized by  $U/t=2(U_{bf}/U_{bb}-1)$ for the coupling constant between nearest neighbors of a Fermi-Hubbard model \cite{Lewenstein04}. Here $t=2J^2/U_{bf}$ is the tunneling energy of the composites, and $J$ is the corresponding energy for single particles. We assume for simplicity equal values of J for both bosons and fermions, but the scheme can be easily generalized to the case $J_b\neq J_f$. The full tight-binding Hamiltonian for composites with chemical potential $\mu$ on a square lattice with $N=L*L$ sites reads
\begin{equation}
H=-\mu\sum_{i=1}^{N}c_{i}^{\dagger}c_{i}-t\sum_{<i,j>}c_{i}^{\dagger}c_{j}-\frac{U}{2}\sum_{<i,j>}c_{i}^{\dagger}c_{i}c_{j}^{\dagger}c_{j}. \label{eq:FermiHubbardH}\end{equation}
In order for this single-band approximation to be valid, the interband gap should be larger than both the thermal and interaction energies. This can easily achieved in typical deep lattice setups.
 Employing an s-wave Feshbach resonance to increase $U_{bf}$ or decrease $U_{bb}$  allows one to reach the large values of $U/t$ needed to obtain, as we will see, bulk gaps of order the hopping $t$ \cite{t}, and therefore (within a mean-field calculation) estimated critical temperatures of order the super-exchange interaction \cite{beyondMF}. 
 Super-exchange effects have been recently observed in a bosonic gas \cite{Trotzky08}, and with the development of novel cooling strategies (see e.g.\cite{coolingStrategies}) such physics seems to be at reach in the near future in lattice mixtures.
 The desired control over the relevant parameters could for example be obtained in a $^{171}$Yb-$^{174}$Yb mixture, where at zero magnetic field one has $U_{bf}/U_{bb}\sim 4$ \cite{Kitagawa08}, or in a $^{6}$Li-$^{7}$Li mixture, where $U/t$ may be tuned employing well-characterized homonuclear and heteronuclear s-wave Feshbach resonances \cite{vKempen04}.
 It should be noticed that in our proposal three-body losses are expected to be severely limited, since we employ only s-wave Feshbach resonances, and in real (virtual) processes sites are never occupied by more than one (two) particle(s).


Following the standard BCS treatment, we introduce the $p$-wave
gap function $\Delta_{ij}=U\langle c_{i}c_{j}\rangle$, which is naturally  
antisymmetric due to the anticommuting properties of the fermionic
operators $c_{i}$. Keeping only quadratic terms in the fluctuations
around the mean-field, the Hamiltonian may be diagonalized by means
of the operators $\gamma_{n}=\sum_{i}u_{n}(i)c_{i}+v_{n}(i)c_{i}^{\dagger}$
$(n=1,\ldots,2N)$, leading to the BdG equations (c.f. \cite{Gurarie07}):
\begin{equation}
\left(\begin{array}{cc}
h_{0} & -\hat{\Delta}^{\dagger}\\
-\hat{\Delta} & -h_{0}\end{array}\right)\left(\begin{array}{c}
u_{n}\\
v_{n}\end{array}\right)=E_{n}\left(\begin{array}{c}
u_{n}\\
v_{n}\end{array}\right).\label{eq:BogolubovH}\end{equation}
Here $h_{0}u_{n}(i)=-\mu u_{n}(i)-t\sum_{\langle j,i\rangle}u_{n}(j)$
is the ideal-gas tight-binding Hamiltonian, while the
pairing operator $\hat{\Delta}$ satisfies $\hat{\Delta}^{\dagger}=-\hat{\Delta}^{*}$,
and acts on the quasiparticle wavefunctions as $\hat{\Delta}u_{n}(i)=\sum_{\langle i,j\rangle}\Delta_{ij}u_{n}(j)$
(the two sums above run over $j$ only). The $2N$ allowed
energies appear in pairs of opposite sign since the Hamiltonian is
symmetric under the exchange of $\{E_{n},\psi_{n}\}$ with $\{-E_{n},\sigma_{1}\psi_{n}^{*})\}$,
where $\sigma_{1}$ is the first Pauli matrix and $\psi_{n}=(u_{n},v_{n})^{T}$.
Anyway, only the $N$ excitations with $E_{n}\geq0$ are physical and will
be considered in the following.
As a consequence of this symmetry, if there exists a solution with $E_0=0$, the associated wavefunction
may be chosen to satisfy the condition $u_{0}=v_{0}^{*}$.
 The self-consistency condition, or
gap-equation, reads
\begin{equation}
\Delta_{ij}=V\langle c_{i}c_{j}\rangle=U\sum_{E_{n}>0}u_{n}^{*}(i)v_{n}(j)\tanh\left(\frac{E_{n}}{2k_{B}T}\right),
\label{eq:gapEq}
\end{equation}
with $k_B$ the Boltzmann constant and $T$ the temperature.
The total number of particles is $N_{{\rm tot}}=\sum_{i}\langle c_{i}^{\dagger}c_{i}\rangle=\sum_{E_{n}>0,i}\left[|v_{n}(i)|^{2}+f(E_{n})(|u_{n}(i)|^{2}-|v_{n}(i)|^{2})\right]$
where $f(E)$ is the Fermi distribution, and the lattice filling is
$F=N_{{\rm tot}}/N$.

On a lattice, the dispersion relation for a homogeneous (vortex-free) chiral $(p_{x}\pm{\rm i}p_{y})$ superfluid with order parameter $\Delta(\mathbf{k})=\Delta_{h}[\sin(k_{x}d)\pm{\rm i}\sin(k_{y}d)]$ reads \cite{Iskin05} \begin{equation}
E(\mathbf{k})=\sqrt{\xi^{2}+|\Delta_{h}|^{2}[\sin^{2}(k_{x}d)+\sin^{2}(k_{y}d)]}\label{eq:dispersionRel}\end{equation}
where $\xi=-\mu-2t[\cos(k_{x}d)+\cos(k_{y}d)]$, and $d$ is the lattice
constant. This dispersion presents gapless points for three values
of $\mu$, i.e. $\mu=-4t,0,4t$. In the absence of interactions, these
values correspond respectively to average fillings $F=0,1/2,1$. When
$\Delta_{h}\neq0$ the dispersion vanishes linearly around the gapless
points, see Fig.~\ref{fig:dispersionRel}, i.e, it coincides with
the one of the Dirac equation.

\begin{figure}
{\tiny(a)}\includegraphics[width=0.5\columnwidth]{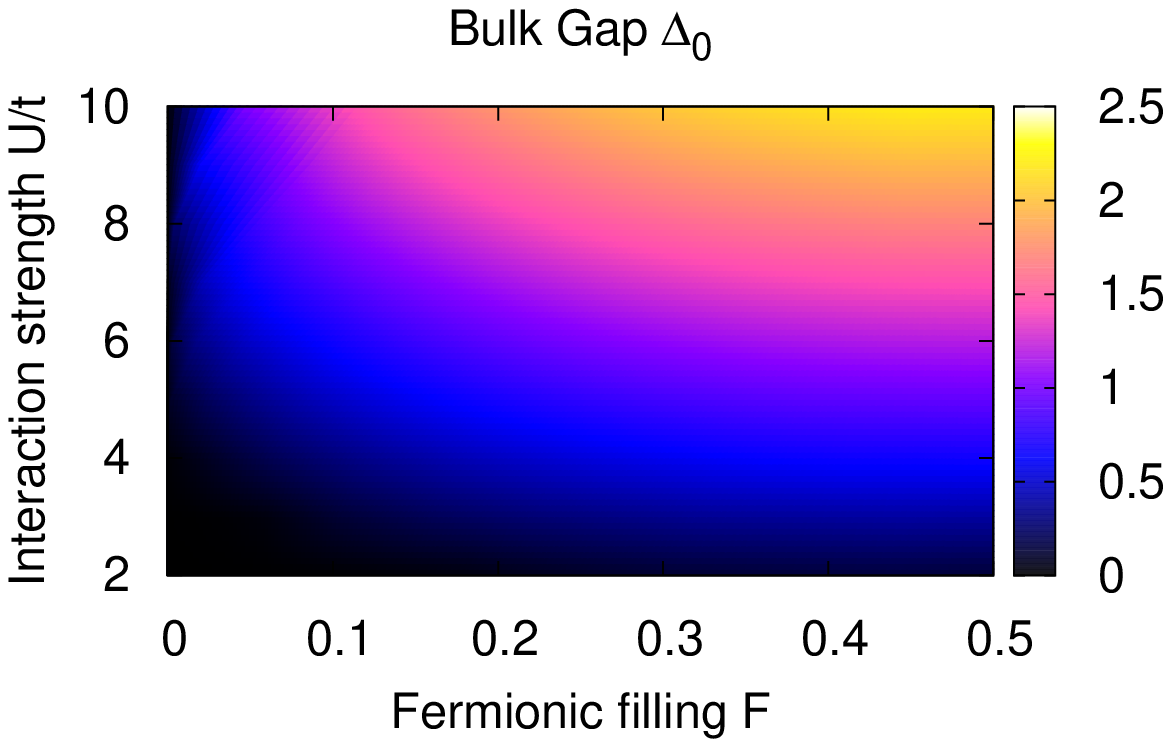}{\tiny(b)}\includegraphics[width=0.5\columnwidth]{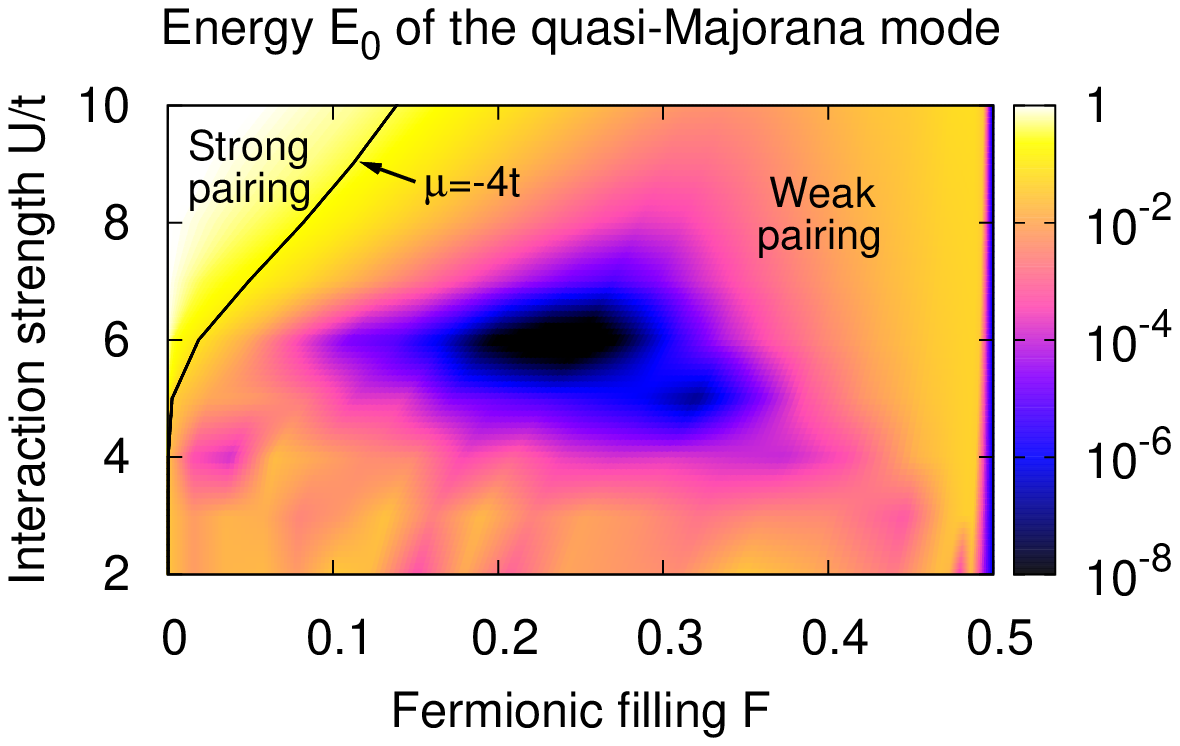}
{\tiny(c)}\includegraphics[width=0.5\columnwidth]{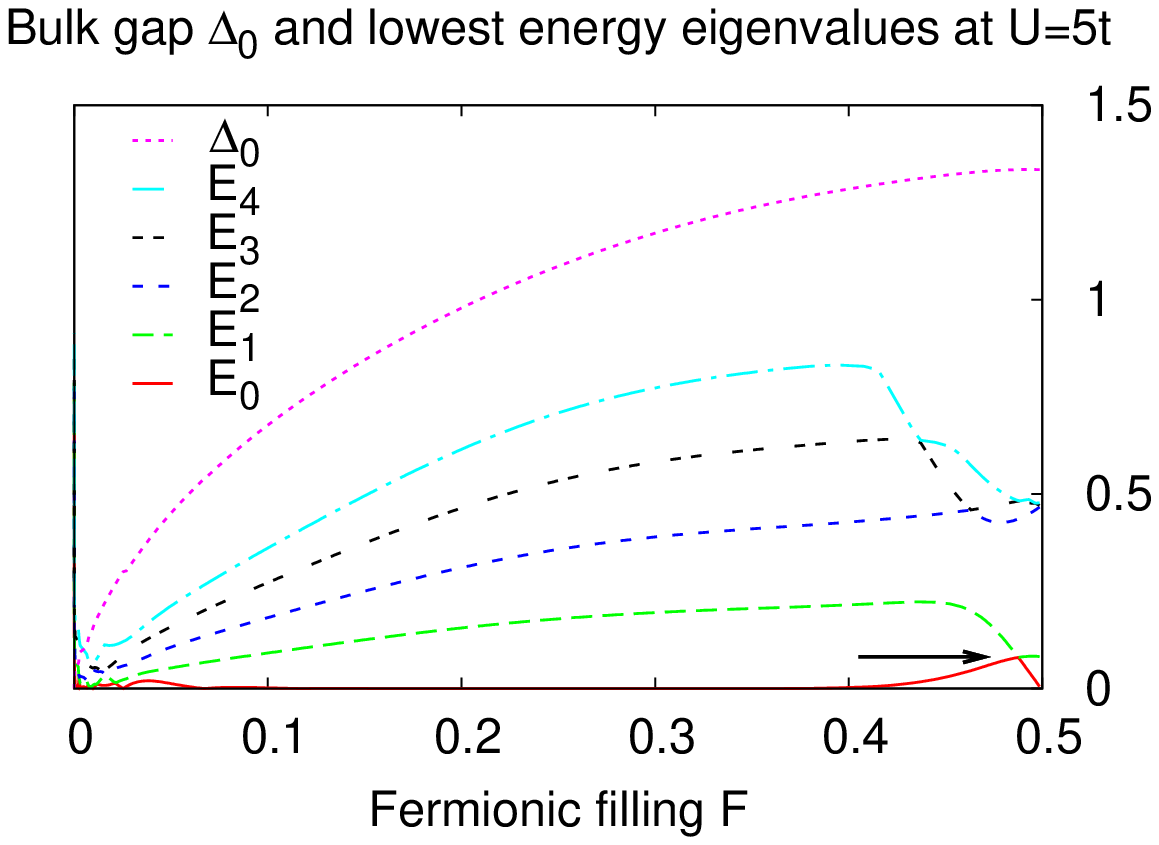}{\tiny(d)}\includegraphics[width=0.5\columnwidth]{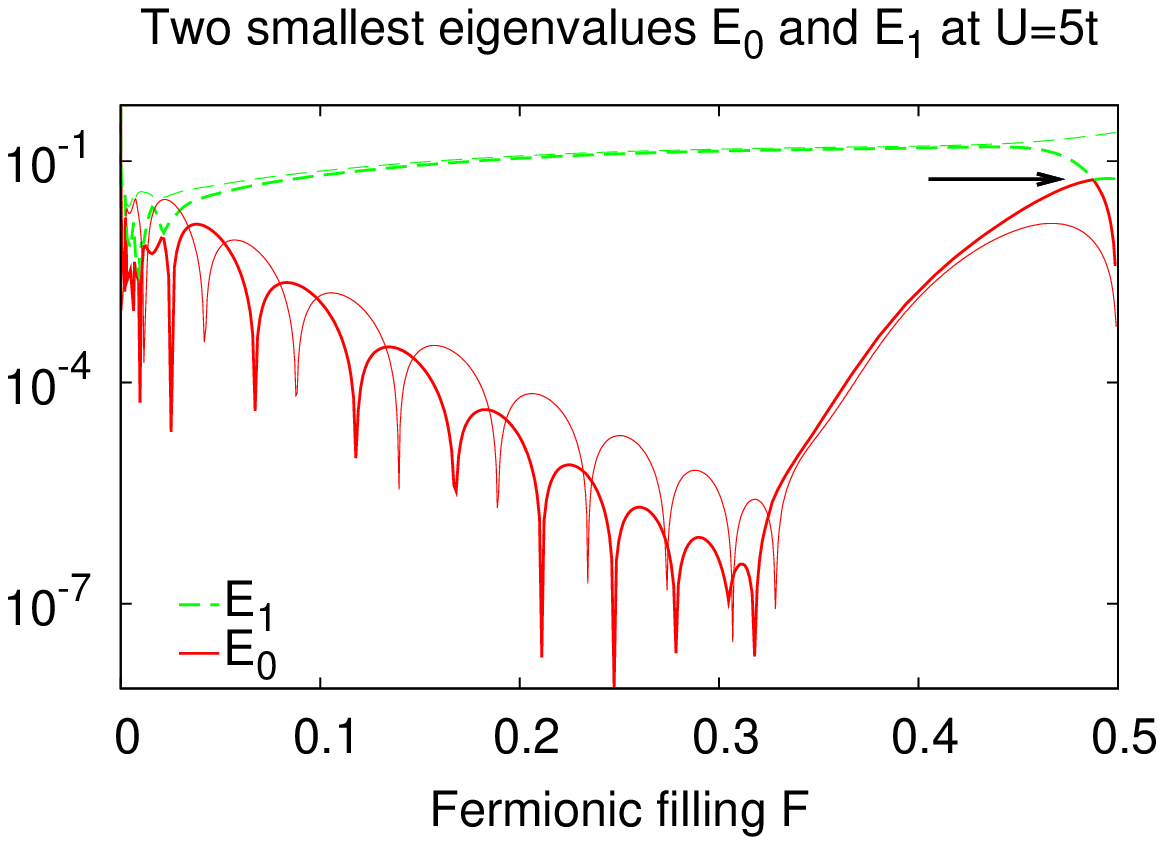}

\caption{Bulk gap $\Delta_{0}$ and lowest energy eigenvalues $E_{i}$
 for the case of a vortex rotating together with the chirality ($w=+1$). Energies in units of $t$. In Fig.\ \ref{fig:Energy}b, the thick line at $\mu=-4t$ marks the TPT from strong 0000 to weak 1000 pairing. The non-trivial TPT from weak 1000 to weak 1110 pairing lies at F=0.5. In the lower figures (where $U=5t)$ the arrow marks the crossing of two eigenvalues causing the abrupt spreading of $\psi_0$ close to the TCP at $F=1/2$. The thin lines in (d) represent the case $w=-1$. 
\label{fig:Energy}}

\end{figure}


Topological modes are known to arise as excitations bound to the core of a vortex in the order parameter.
 To model the vortex, we use open
boundary conditions and solve self-consistently Eqs.~(\ref{eq:BogolubovH}-\ref{eq:gapEq})
by taking the initial Ansatz for the gap $\Delta_{ij}=\mbox{\ensuremath{\chi}}[f_{i}\mathrm{e}^{\mathrm{i}w\theta_{i}}+f_{j}\mathrm{e}^{\mathrm{i}w\theta_{j}}]/2$.
Here $f_{i}=r_{i}/\sqrt{r_{i}^{2}+2\sigma^{2}}$ is the amplitude
of the gap around a vortex with core size $\sigma$,
and $r_{i}$ and $\theta_{i}$ are the polar coordinates of site $i$
with respect to the core. The chirality factor $\chi$ for
$p_{x}+{\rm i}p_{y}$ pairing takes the values $\chi=\pm1$ when
$j=i\pm\hat{\mathbf{x}}$, and $\chi=\pm{\rm i}$ when $j=i\pm\hat{\mathbf{y}}$.
Finally, $w=\pm1$ is the direction of rotation of the vortex with
respect to the chirality. The results presented here are obtained	
on a lattice with $25*25$ sites, which is experimentally feasible, and large enough to obtain
a satisfactory qualitative description of larger lattices \cite{fotka}.

In the absence of a confining potential, the physics in
the lattice exhibits  particle-hole symmetry around the half-filling point $\mu=0$. For $\mu<0$ ($\mu>0$), one has $0<F<1/2$ ($1/2<F<1$), and the superfluid
may be identified with a condensate of pairs of particles (holes). The gap function and the energy of the eigenstates
depend only on the modulus of $\mu$, but the wavefunctions $u$ and $v$ depend also on its sign, since the two phases 1000 and 1110 are topologically distinguishable \cite{Kou09}.

In the continuum, zero energy modes appear when the chemical potential becomes positive, the so-called weak pairing phase. Correspondingly, in a lattice the weak pairing phases are found for chemical potentials in the range $-4t<\mu<0$ and $0<\mu<4t$ for particles and holes, respectively.
As shown in Fig.~\ref{fig:Energy}, at intermediate fillings and for sufficiently strong interactions $(t\lesssim U\lesssim10t)$ the T=0 bulk gap $\Delta_0$ (defined as the modulus of $\Delta_{\langle i,j \rangle}$ away from the vortex core) rises to values $\sim t$ indicating large pairing correlations, and
the low energy  spectrum of the BdG Hamiltonian becomes linear,
$|E_{n}|\sim n\omega_{0}$, where $n=0,1,2,\ldots$ \cite{Kopnin91}. In particular, there exists a quasi-Majorana fermion excitation $\psi_{0}$ with $u_0\simeq v_0^*$ and an extremely small eigenvalue, $E_{0}\ll\Delta_{0}$. For very strong attraction ($U\gtrsim10t$) the vortex core size shrinks below a lattice constant, and the lowest energy excitation loses its topological quasi-Majorana character. Moreover, too large values of $U/t$ should be avoided since they lead to phase separation \cite{Wang05,Titvinidze08}.
As intuitively expected from s-wave BCS theory, we have checked that a quasi-Majorana mode is present as long as the temperature is smaller than $\Delta_0$, the bulk gap at $T=0$.

\begin{figure}
\includegraphics[width=0.5\columnwidth]{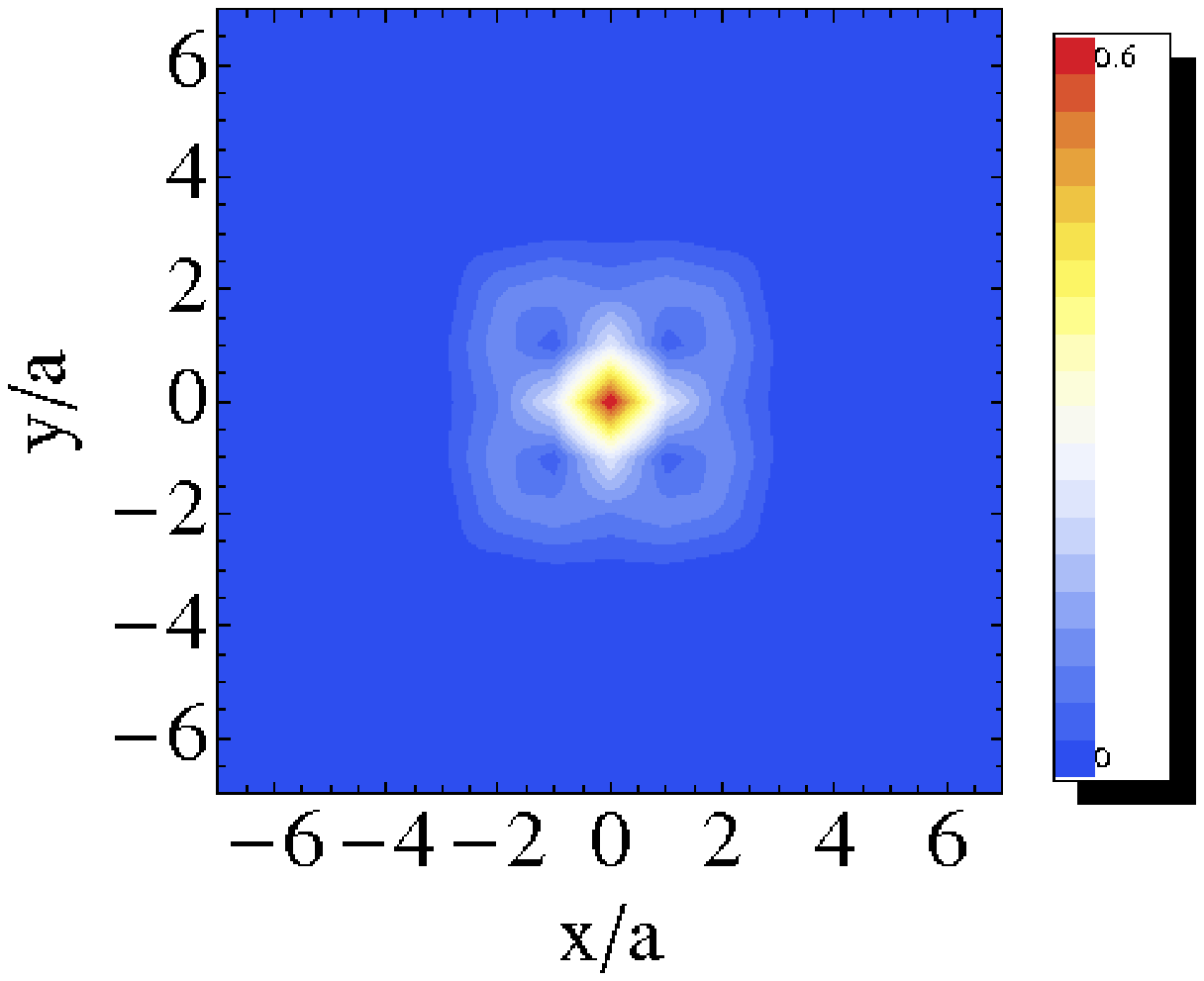}\includegraphics[width=0.5\columnwidth]{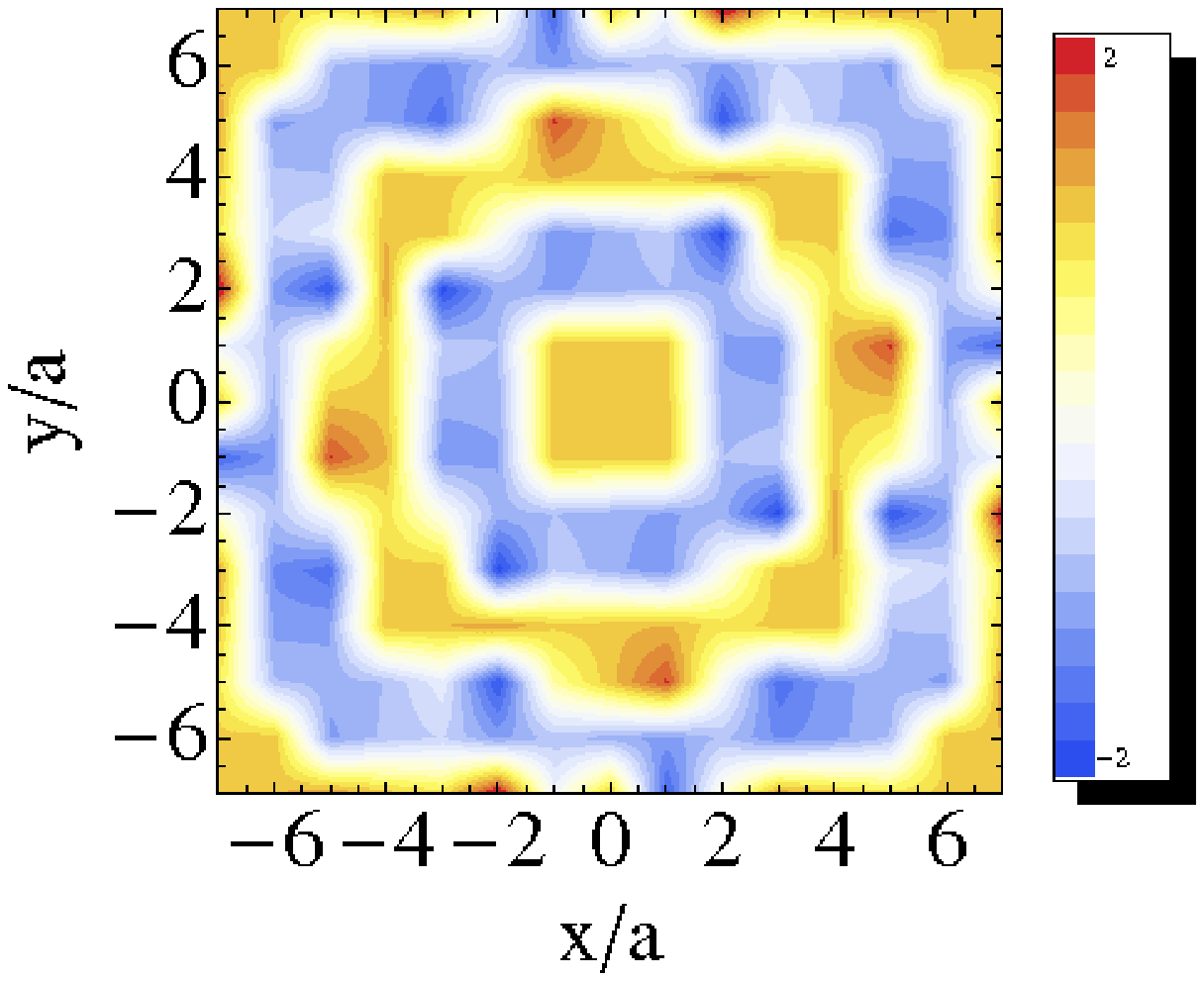}
\caption{Wavefunction $u_0\simeq v_0^*$ at $F\sim1/4$, $w=-1$, and $U=5t$. Left: the
modulus is exponentially localized around the vortex core. Right:
the phase plot evidences the shape of the nodes adapting to the lattice
structure. \label{fig:quartFillWaveFunction}}

\end{figure}

In Fig.~\ref{fig:quartFillWaveFunction} we show a typical $\psi_{0}$ in the weak pairing phase; similarly to
the continuum zero-modes \cite{Read00,Tewari07,Gurarie07}, it
exhibits  oscillations which are exponentially damped away from the vortex core.
 When the vortex rotates in the (opposite to the) direction of the chirality, i.e., $w=+1$ ($w=-1$),
 $|\psi_{0}|$ has a node (a maximum) at the vortex core. 
The lattice effects are clearly reflected in four typical features: 
the energy of quasi-Majorana mode $E_{0}$ exhibits oscillations as a function of  $F$ (see Fig.~\ref{fig:Energy}d), corresponding to the appearance of new nodes within the
finite extension of the lattice; the nodal structure
of the wavefunctions is aligned along the lattice axes to minimize
energy; for the level spacing we find $\omega_{0}\sim\Delta_{0}$, while in the continuum one has $\omega_{0}\sim\Delta_{0}^{2}/\epsilon_{F}$
with $\epsilon_{F}$ the Fermi energy; finally, at half-filling the momentum distribution $\psi_{0}(\mathbf{k})$
is peaked at the minima (Dirac points) of the dispersion relation of Fig.~\ref{fig:dispersionRel}(center).

\begin{figure}
\includegraphics[width=0.2\columnwidth]{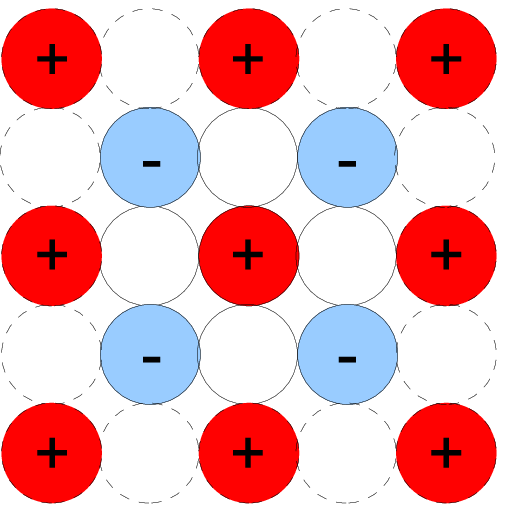}\includegraphics[width=0.4\columnwidth]{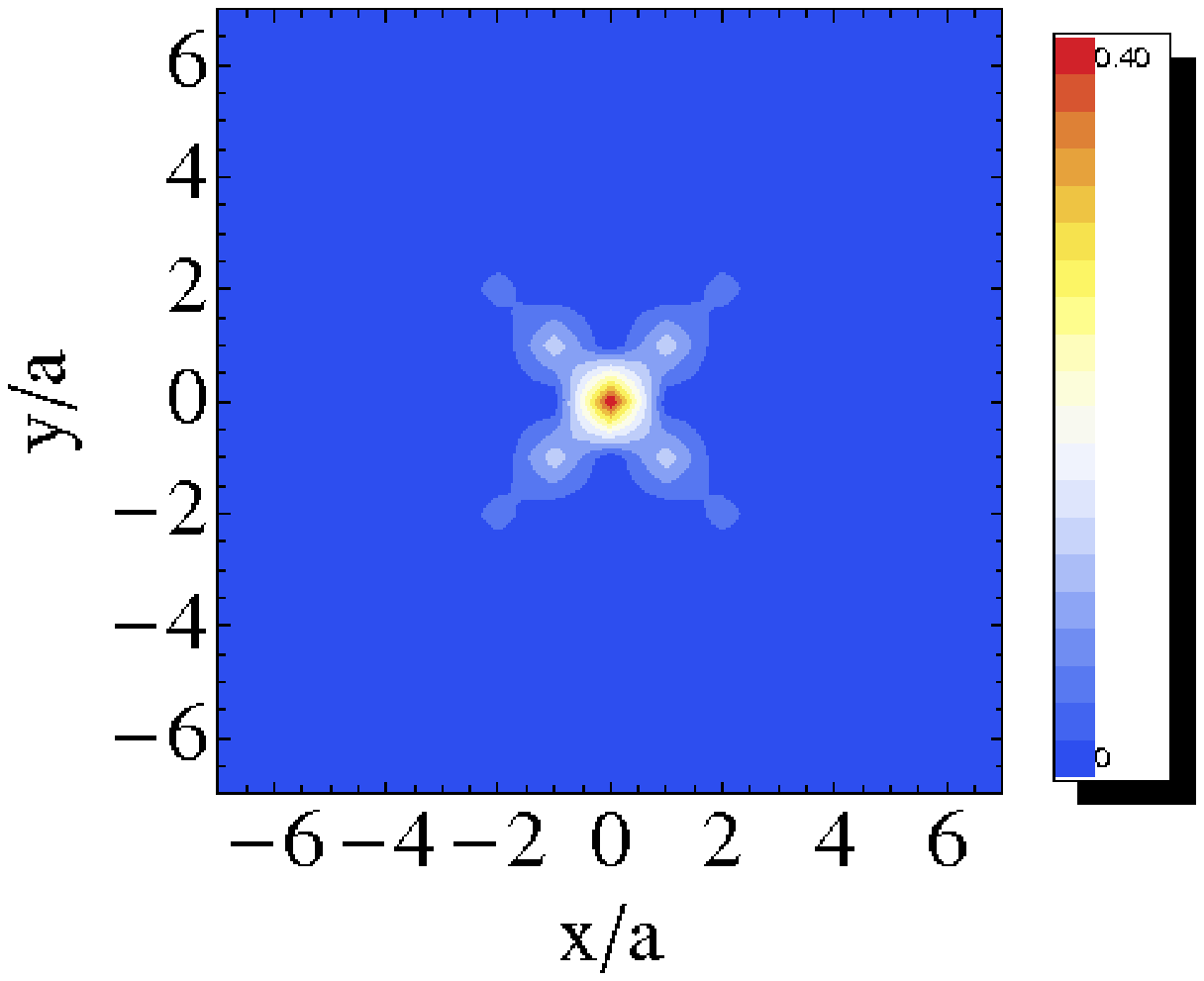}\includegraphics[width=0.4\columnwidth]{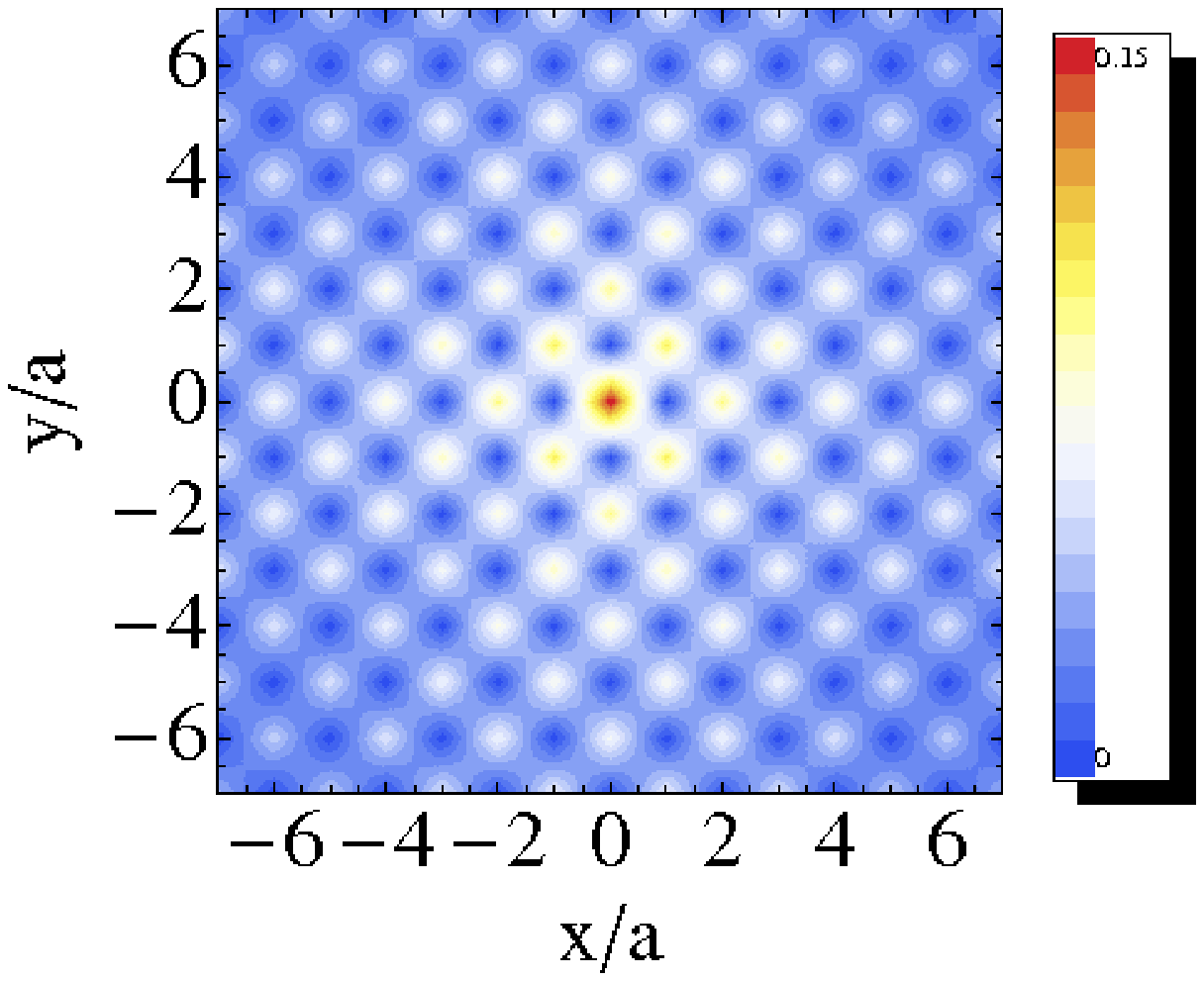}
\caption{Wavefunction $u_0\simeq v_0^*$ at half-filling $(\mu=0)$. Left: sketch of the d-wave checkerboard structure (+ and - refer to the sign
of the ${\rm Re}(u_{0})$, while ${\rm Im}(u_{0})=0$).
 Center and right: modulus of $u_0$ for $w=-1$ and $w=+1$, respectively.
\label{fig:halfFillWaveFunction}}
\end{figure}

The lattice effects are most strongly noticeable at half-filling
($\mu=0$), where the lattice spacing gets to coincide with half the
period of the lowest energy excitation oscillations. Already in the homogeneous, vortex free case, 
there exist two zero-energy ``stripe modes'', since the dispersion vanishes at the gapless points $\mathbf{k}=(0,\pm\pi)$
and $\mathbf{k}=(\pm\pi,0)$, see Eq.~(\ref{eq:dispersionRel}) and
Fig.~\ref{fig:dispersionRel}(center), and these modes may combine to form a checkerboard d-wave pattern.
 In presence of a vortex Eq.~(\ref{eq:dispersionRel}) does not hold, and the situation is more complex. 
The system minimizes its energy by arranging itself in a d-wave checkerboard pattern, the vortex core being occupied, see Fig.~\ref{fig:halfFillWaveFunction}.
With open boundary conditions, however, this pattern is commensurate with the
lattice only when $L$ (the number of sites per lattice side) is odd:
in this case, the corresponding mode  has  a vanishingly small eigenvalue \cite{Kou09}.
If $L$ is even, this zero is lifted and the spectrum
contains two identical lowest eigenvalues. This situation is reminiscent of the spectrum of a non-interacting chain, where the presence of a zero eigenvalue depends on its even or odd length. Lattice effects at $\mu=0$ also destroy the linearity of the low energy part of the spectrum, see Fig.~\ref{fig:Energy}c.

For odd $L$, $\psi_{0}$ strongly depends on the direction of rotation
of the vortex with respect to the chirality, as shown in Fig.~\ref{fig:halfFillWaveFunction}. When the two counter-rotate
($w=-1$), the even symmetry of the wavefunction is compatible with
the requirement of d-wave symmetry at half filling. As $F$ approaches 1/2, $\psi_{0}$
smoothly evolves towards a wavefunction tightly localized around the core.
When instead $w=1$, as discussed above, away from half-filling the lowest energy excitation has a node in the vortex core. This is not compatible with the checkerboard
structure minimizing the energy at half-filling.
 As a consequence, approaching the Topological Critical Point (TCP) $\mu=0$ the symmetry of the lowest energy excitation suddenly changes, and $\psi_0$ spreads over the whole lattice. The sudden spreading of $\psi_0$ is signalled by the sharp crossing of the two lowest eigenvalues, as marked by an arrow in Fig.~\ref{fig:Energy}c-d.
On a finite lattice, the detection of this extended state would therefore be a clear experimental signature of the vicinity of the TCP.

The general features discussed above persist even
in the presence of an external harmonic confinement, provided this varies sufficiently
slowly in the central part of the lattice. In particular, we find that the low-energy
section of the (positive) spectrum in the weak pairing phase is still
approximately linear and admits an eigenvalue $E_{0}\ll\Delta_{0}$.
Nonetheless, the confinement destroys the aforementioned particle-hole
symmetry, and the spatial inhomogeneity makes
it difficult to observe the formation of the checkerboard state around half-filling.
To avoid this problem, optical potentials formed by higher order Laguerre-Gauss beams, or by ``light sheets'', could be used; they yield potentials of the form $\propto\left(x^{2\nu}+y^{2\nu}\right)$, characterized by a rather uniform central region and steep confining walls.
 Simulations for the experimentally feasible case $\nu=2$, $w=1$, and  $F\simeq 0.22$, lead to $\mu=0$, and exhibit quite nicely 
a checkerboard quasi-zero mode with significant extension. 
 This structure
could be detected experimentally via spatially-resolved polarization
spectroscopy \cite{Eckert08}, while Majorana modes may be observed via radio-frequency absorption \cite{Grosfeld07}.

In conclusion, we have proposed a novel scheme to realize stable p-wave superfluids in optical lattices, and we have shown that even in presence of a relatively small lattice, the lowest energy excitation preserves its topological character.
Optimal filling factors and interaction strengths to obtain Majorana modes are $F\sim1/4$ (or equivalently $F\sim3/4$) and $U/t\sim5$, where the ratio $E_0/\Delta_0$ is smallest. At the critical point $\mu=0$, the system still possesses a zero mode, but the linear structure of the low energy part of the spectrum is lost. Here, the system favors the formation of checkerboard order. If the vortex rotates in the same direction as the chirality, the vicinity of the topological phase transition is signalled by the abrupt spreading of the lowest energy excitation wavefunction. These results are robust for temperatures below the super-exchange interaction and even in presence of an external confinement. 

We wish to thank F. Cucchietti, O. Dutta, A. Eckardt, V. Gurarie, and A. Kubasiak for interesting discussions. We acknowledge Spanish MEC projects
FIS2008-00784, FIS2008-01236 and QOIT (Consolider Ingenio 2010), ESF/MEC
project FERMIX (FIS2007-29996-E), EU Integrated Project  SCALA, EU
STREP project NAMEQUAM, ERC Advanced Grant QUAGATUA, and Alexander von Humboldt Found.\ Senior Research Prize.


\begin{thebibliography}{20}
\bibitem{beyondMF} Beyond mean-field corrections should be taken into account in the thermodynamic limit where phase fluctuations with vanishing wavevector are known to destroy long-range order, but finite systems have a natural infrared cutoff which blocks this effect.
\bibitem{fotka} We considered lattice sizes up to $\sim50*50$, without observing qualitative changes. Our findings are also independent from the vortex amplitude Ansatz $f_i$.
\end{thebibliography}
\end{document}